\documentclass[10pt,aps,prb,twocolumn,superscriptaddress,showpacs]{revtex4-1}

\usepackage[pdftex]{graphicx}
\usepackage{hyperref}
\usepackage{amssymb}
\usepackage{amsmath}
\usepackage{xspace}
\usepackage{color}

\begin{document} 

\title{Ising tricriticality in the  extended
Hubbard model with bond dimerization}

\author{Satoshi Ejima}
\affiliation{Institut f\"ur Physik, Ernst-Moritz-Arndt-Universit\"at
Greifswald, 17489 Greifswald, Germany}

\author{Fabian H. L. Essler}
\affiliation{The Rudolf Peierls Centre for Theoretical Physics, 
Oxford University, Oxford OX1 3NP, United Kingdom}

\author{Florian Lange}
\affiliation{Institut f\"ur Physik, Ernst-Moritz-Arndt-Universit\"at
Greifswald, 17489 Greifswald, Germany}
\affiliation{
Computational Condensed Matter Physics Laboratory, RIKEN, Wako, Saitama 351-0198, Japan}

\author{Holger Fehske}
\affiliation{Institut f\"ur Physik, Ernst-Moritz-Arndt-Universit\"at
Greifswald, 17489 Greifswald, Germany}

\date{\today}

\begin{abstract}
We explore the quantum phase transition between Peierls and 
charge-density-wave insulating states in the one-dimensional,
half-filled, extended Hubbard model with explicit bond dimerization. 
We show that the critical line of the continuous Ising transition
terminates at a tricritical point, belonging to the universality class 
of the tricritical Ising model with central charge $c=7/10$. 
Above this point, the quantum phase transition
becomes first order. Employing a numerical matrix-product-state based 
(infinite) density-matrix renormalization group method we determine the
ground-state phase diagram, the spin and two-particle charge excitations
gaps, and the entanglement properties of the model with high precision. 
Performing a bosonization analysis we can derive a field description of 
the transition region in terms of a triple sine-Gordon model. 
This allows us to derive field theory predictions for the power-law 
(exponential) decay of the density-density (spin-spin) and bond-order-wave
correlation functions, which are found to be in excellent agreement 
with our numerical results.
\end{abstract}

\maketitle

\section{Introduction}
Ising tricriticality emerges at the end point of a continuous line of Ising
quantum phase transitions, above which a first-order transition
occurs. In 1+1 dimensions, it is described by a conformal field theory
(CFT) and more precisely the second minimal model of central charge
$c=7/10$.~\cite{FQS84,FQS85} Interestingly, the tricritical Ising
model (TIM) exhibits space-time supersymmetry. Until recently, there
were only a few known condensed matter realizations of the TIM
such as the Blume-Capel model~\cite{Blume66,Capel66,ADKS85}
or the so-called golden chain with Fibonacci anions.~\cite{FTLTKWF07}
In the last couple of years, other realizations were found in 
lattice models with interacting Majorana fermions,~\cite{RZFA15,ZF16}
and in an extended Hubbard model (EHM) with on-site ($U$) and
nearest-neighbor ($V$) Coulomb interactions, in a case where an (somewhat
artificial) alternating ferromagnetic spin interaction ($J$) was
added.~\cite{LEF15} In this model, the $U$ and $V$ terms induce
respectively fluctuating spin-density-wave (SDW) and
charge-density-wave (CDW) order. The $J$ term promotes the formation
of spin-1 moments (out of two spins on neighboring sites) and the
build-up of a symmetry-protected topological (SPT)
state,~\cite{PTBO10} in close analogy to the spin-1 $XXZ$ chain.
As a result, the SDW gives way to a Haldane insulator (HI), and a
quantum phase transition takes place between the HI and the CDW when
$V$ increases. If this HI-CDW Ising transition line meets a
first-order transition line, a  tricritical Ising point appears. 

Another, perhaps more realistic, model system, 
attracting a lot of attention, is the half-filled EHM with  
explicit bond dimerization.~\cite{TF04,BEG06} Here the formation of an
SPT phase might be triggered by the Peierls instability. 
Indeed, the ground-state phase diagram, obtained within a (perturbative)
weak-coupling approach,~\cite{TF04}  contains besides the CDW a
bond-dimerized phase. In order to distinguish this phase from the
bond-order-wave (BOW) phase in the EHM,~\cite{Na99,Na2000} which
arises as a result of spontaneous symmetry breaking, we will call it
a \emph{Peierls insulator} (PI) in the following. The quantum phase transition
line between the insulating CDW and PI phases belongs to the universality class
of the two-dimensional Ising model,~\cite{TF04,BEG06} and has been
argued to terminate in a tricritical point, where the phase transition
changes from continuous to first order.  
The existence and universality class of the tricritical point
is an open question however. To address this issue, not only a
numerical study should be possible (e.g., along the lines of
Ref.~[\onlinecite{LEF15}]), but also a field theoretical analysis,
based on the results of Ref.~[\onlinecite{BEG06}].   

The aim of the present work is to establish the tricritical Ising
universality class at the tricritical  point on the PI-CDW transition
line of the half-filled EHM  with staggered bond dimerization, using
both a matrix-product-state (MPS) based numerical density-matrix
renormalization group (DMRG) technique~\cite{Wh92} and a bosonization
approach \cite{GNT99,EFGKK05} combined with a field
theoretical analysis.  

The outline of this paper is as follows. 
In Sec.~\ref{sec:model}, we introduce and motivate the model Hamiltonian 
under investigation.
Section~\ref{sec:dmrg} presents our DMRG results,  in particular the
ground-state phase diagram,  the excitation gaps, and the entanglement entropy. 
Section~\ref{sec:FT} describes the field theoretical approach and makes predictions for
the quantum critical line, as well as for the density-density, spin-spin,  
and bond-order-wave correlations (see also Appendix),
which can be used to analyze our numerical data. We conclude in Sec.~\ref{conclusion}.

\section{Model}
\label{sec:model}
The Hamiltonian of the EHM is defined as
\begin{eqnarray}
\hat{H}_{\textrm{EHM}} &=&
 -t \sum_{j\sigma}
 (\hat{c}^\dagger_{j\sigma}\hat{c}_{j+1\sigma}^{\phantom{\dagger}} 
 + {\rm H.c.}) 
 \nonumber\\
 &&+U \sum_{j}\left(\hat{n}_{j\uparrow}-\frac{1}{2}\right)
                \left(\hat{n}_{j\downarrow}-\frac{1}{2}\right)
 \nonumber\\
 && + V \sum_{j} (\hat{n}_{j}-1)
  (\hat{n}_{j+1}-1) \,,
  \label{EHM}
\end{eqnarray}
where $\hat{c}^\dagger_{j\sigma}$ ($\hat{c}^{\phantom{}}_{j\sigma}$)
creates (annihilates) an electron with spin 
$\sigma=\uparrow,\downarrow$ in a Wannier orbital centered 
around site~$j$,
$\hat{n}_{j\sigma}=\hat{c}^\dagger_{j\sigma}\hat{c}^{\phantom{}}_{j\sigma}$,
and $\hat{n}_{j}=\hat{n}_{j\uparrow}+\hat{n}_{j\downarrow}$. 
For $V=0$, the ground state has fluctuating SDW order
(there is no long-range order, but the dominant correlations are of
SDW type) with gapless spin and gapped charge excitations 
$\forall U>0$.~\cite{EFGKK05} 
In the regime $V/U\lesssim1/2$, the ground state remains
a SDW, but acquires 2$k_{\textrm  F}$-CDW order when
$V/U\gtrsim1/2$. The SDW and CDW phases are separated by a narrow
BOW phase below the critical end
point.\cite{SSC02,SBC04,Zhang04,TTC06,EN07}
The BOW phase exhibits spontaneous breaking of translational symmetry
and is characterized by a staggered modulation of the kinetic energy
density. 
Adding a staggered ferromagnetic spin interaction,   
 $\hat{H}_{J}=J\sum_{j=1}^{L/2}
  \hat{\mbox{\boldmath $S$}}_{2j-1}\hat{\mbox{\boldmath $S$}}_{2j} $
with $\hat{\mbox{\boldmath $S$}}_j=
   (1/2)\sum_{\sigma\sigma^\prime}
      \hat{c}_{j\sigma^{\phantom{\prime}}}^\dagger 
      \mbox{\boldmath $\sigma$}_{\sigma\sigma^\prime}^{\phantom{\dagger}}
      \hat{c}_{j\sigma^\prime}^{\phantom{\dagger}}$, to the 1D EHM, 
the alternating spin exchange tends to form spin-1 moments with the
result that the SPT  HI~\cite{PTBO10} replaces the Mott insulating 
and BOW states of the EHM at small $V/U$.~\cite{LEF15} 

In the following, we ask whether a similar scenario holds
for the half-filled  EHM  with staggered bond dimerization:
\begin{equation}
 \hat{H} = \hat{H}_{\textrm{EHM}}+\hat{H}_{\delta}\;,
  \label{EPHM}
\end{equation}
\begin{equation}
 \hat{H}_{\delta}= -t\sum_{j\sigma}\delta(-1)^j
 (\hat{c}^\dagger_{j\sigma}\hat{c}_{j+1\sigma}^{\phantom{\dagger}}
 + {\rm H.c.})\, .
\end{equation}
It was previously shown that in the large-$U$ limit the low-lying
excitations of (\ref{EPHM}) are chargeless spin triplet 
and spin singlet
excitations,~\cite{GNT99,Gi03,GBSTK97,NF81,Ts92,US96,ETD97} 
whose dynamics is described by a spin-Peierls Hamiltonian.

For finite $U$, the Tomonaga-Luttinger liquid para\-meters 
have been determined at and near commensurate band fillings,~\cite{EGN06} 
by means of DMRG calculations.
In the weak electron-electron interaction regime,
perturbative\cite{GGR05,DM05} and renormalization
group\cite{SS02,TO02,TF04} approaches determined that the system
realizes PI and CDW phases at half-filling.
Exploiting DMRG and field theory, it was shown that the transition 
between these two phases belongs to the universality
class of the two-dimensional Ising model.~\cite{TF04,BEG06}

\section{DMRG treatment}
\label{sec:dmrg}
In this section, we examine the ground-state properties 
of the 1D lattice Hamiltonian~\eqref{EPHM} with a high accuracy
by means of the MPS-based infinite DMRG (iDMRG)
technique.~\cite{Mc08,Sch11} 
The method works directly in the thermodynamic limit. 
The PI and CDW boundaries are characterized by various 
excitation gaps obtained by DMRG combined with the infinite MPS representation 
on the boundaries, see previous work by some of the
authors.~\cite{LEF15}  When tracing the central charge along the
PI-CDW transition line, we use DMRG for finite systems with periodic
boundary conditions (PBC).

\subsection{Phase diagram}
\begin{figure}[tb]
 \includegraphics[width=\columnwidth,clip]{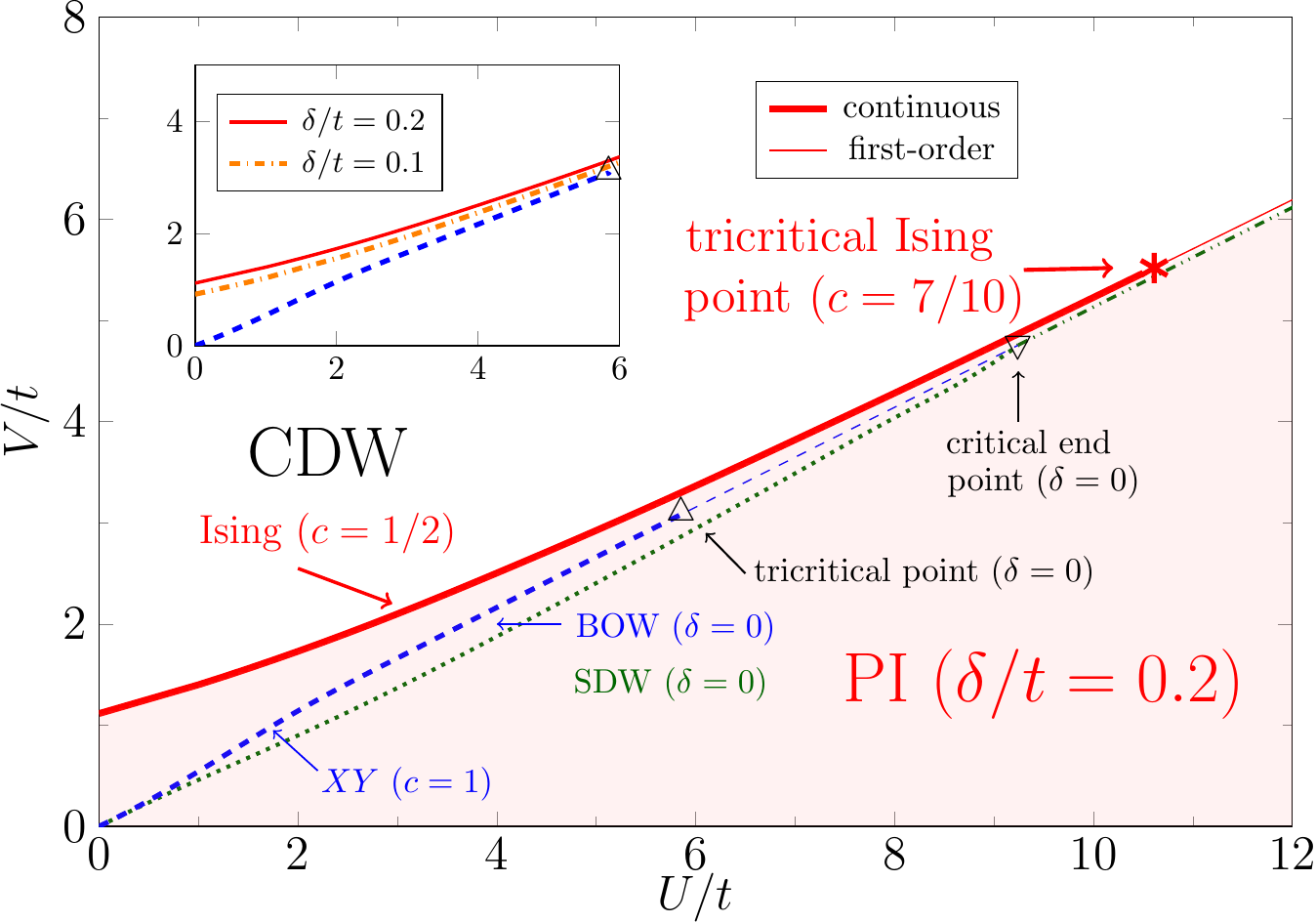} 
 \caption{(Color online) iDMRG ground-state phase diagram of 
 the 1D EHM with bond dimerization~\eqref{EPHM}. 
 The red solid line gives the PI-CDW phase boundaries 
 for $\delta/t=0.2$. The quantum phase transition is continuous (first order) 
 below (above) the tricritical Ising point $[U_{\rm t}, V_{\rm t}]$ marked by 
 the asterisk.  For comparison results for the BOW-CDW (blue dashed line), 
 SDW-BOW (green dotted line), and SDW-CDW (green dashed-dotted 
 line) transitions of the pure EHM ($\delta=0$)
 were included.~\cite{EN07}
 (Inset) PI-CDW transition for $\delta/t=0.1$ and $0.2$ 
 in the weak-coupling regime. As expected, decreasing $\delta/t$, the transition lines come 
 closer to BOW-CDW transition line of the pure EHM.
 }
 \label{pd}
\end{figure}

According to weak-coupling renormalization-group results,~\cite{TF04}
a bond alternation $\delta$  changes the universality class of the BOW-CDW transition
in the EHM  from Gaussian- to Ising-type. The Ising
criticality has been confirmed by DMRG computations.~\cite{BEG06}

Figure~\ref{pd} presents the complete ground-state phase diagram of 
the EHM with bond dimerization, as obtained by the 
iDMRG technique. The phase boundaries for the pure EHM are also
included (blue and green lines). The dimerized PI phase replaces
entirely the SDW and BOW  states of the EHM. The PI state has the
lowest energy also in the weak-coupling regime, and even at
$U/t=0$. This finding confirms previous weak-coupling renormalization
group results.~\cite{TF04} In the intermediate-to-strong coupling
regime, the PI-CDW transition line converges to those of the
BOW/SDW-CDW transition for  the pure EHM. The transition is continuous
up to the tricritical Ising point $[U_{\rm t}, V_{\rm t}](\delta)$,
which converges naturally to the tricritical point of the EHM when
$\delta\to0$. Above $[U_{\rm t}, V_{\rm t}]$, the PI-CDW transition
becomes first order. At very large $U/t$, the phase boundaries of the
PI/SDW-CDW transitions are almost indistinguishable.

We now characterize the different ground states of the
model~\eqref{EPHM} in some more detail.  
Since the dimerized PI state can be considered as an SPT state, 
the entanglement spectrum plays an important role in our analysis.
The so-called entanglement spectrum $\epsilon_\alpha$ can be extracted
from the singular value 
decomposition.~\cite{LEF15} Dividing our system into two subblocks, 
$\cal{H}=\cal{H}_{\textrm L}\otimes\cal{H}_{\textrm R}$, and 
considering the reduced density matrix 
$\rho_{\textrm L}=\mathrm{Tr}_{\textrm R}[\rho]$,
the entanglement spectra are given by the singular values
$\lambda_\alpha$ of $\rho_{\textrm L}$ as 
$\epsilon_\alpha=-2\ln\lambda_\alpha$. Moreover, the correlation length
$\xi_\chi$ can be determined from the second largest eigenvalue of the 
transfer matrix for some bond dimension $\chi$ used in the iDMRG
simulation.~\cite{Mc08,Sch11}
While the physical correlation length diverges at the critical point,
$\xi_\chi$ stays finite, as a consequence of working with a finite
bond dimension $\chi$. Because of  $\xi_\chi$'s rapid increase with $\chi$ near the critical
point, $\xi_\chi$ can be used nevertheless to determine the phase transition.  
We performed iDMRG simulations with $\chi$ up to 400, so that the effective correlation 
length at criticality is less or at most equal 300.

\begin{figure}[tb]
 \includegraphics[width=\columnwidth,clip]{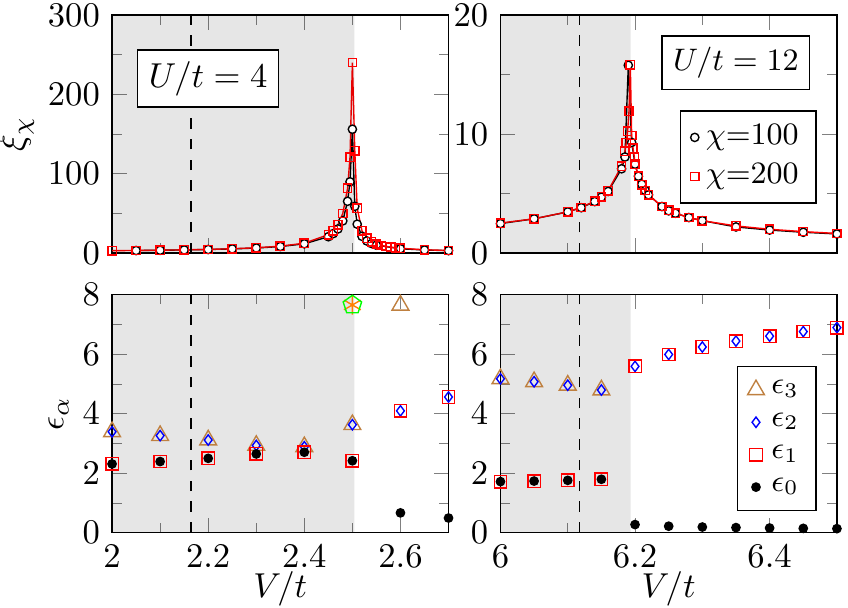} 
 \caption{(Color online) Correlation length $\xi_\chi$ (top)
 and entanglement spectrum $\epsilon_\alpha$ (bottom) as a function of $V/t$
 for $U/t=4$ (left) and $U/t=12$ (right), 
 where $\delta/t=0.2$. Data are obtained by iDMRG.
 Dashed lines give the BOW-CDW (SDW-CDW) transition for $U/t=4$ 
 ($U/t=12$) in the EHM.~\cite{EN07}
 }
 \label{xi-es}
\end{figure}

Figure~\ref{xi-es} gives $\xi_\chi$ and $\epsilon_\alpha$ as functions of $V/t$ 
for fixed $\delta/t=0.2$, at two characteristic  $U/t$ values. 
In the weak-to-intermediate coupling regime, $U/t=4$,
we find a distinct peak in the correlation length 
at $V_{\rm c}/t\simeq2.504$, which increases rapidly as $\chi$ grows
from 100 to 200, indicating the divergence of the correlation length
$\xi_\chi\to\infty$ as $\chi\to\infty$, i.e., 
a quantum phase transition (of Ising type, as will be shown in Sec.~\ref{sec-ent}). 
In contrast, at strong coupling $U/t=12$, the peak
height stays almost constant at $V_{\rm c}/t\simeq6.194$ when $\chi$
is enhanced. 
Decreasing the magnitude of $\delta/t$, the transition points will
approach those of the pure EHM, e.g., for $\delta/t=0.1$ and $U/t=4$
we find $V_{\rm c}/t\simeq2.372$, with a simultaneous reduction of  
the $\xi_\chi$'s peak heights. Most notably, the entanglement spectra 
of the dimerized SPT phase exhibits a distinguishing double degeneracy 
in the lowest entanglement level;~\cite{PTBO10} for $V>V_{\rm c}$, 
in the CDW phase, this level is nondegenerate.

\subsection{Excitation gaps}

\begin{figure}[t]
 \includegraphics[width=0.8\columnwidth]{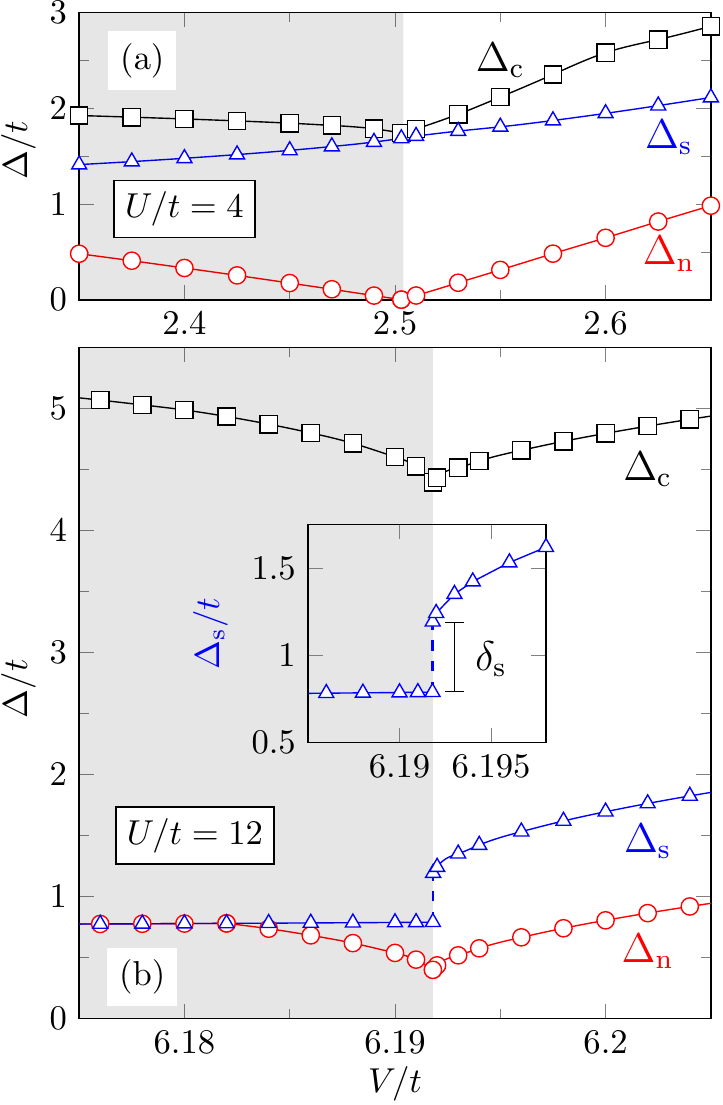}
 \caption{(Color online) Charge ($\Delta_{\rm c}$), spin 
 ($\Delta_{\rm s}$) and neutral ($\Delta_{\rm n}$) gaps in dependence on
 $V/t$ for (a) $U/t=4$ and (b) $U/t=12$.  Again, $\delta/t=0.2$.
 The dimerized PI (CDW) phase is marked in gray (white). Note the jump
 of the spin gap, 
 $\delta_{\rm s}\equiv \Delta_{\rm s} (V_{\rm c}^+) - \Delta_{\rm s} (V_{\rm c}^-)$,
 at $V_{\rm c}/t$.
 }
 \label{gaps}
\end{figure}
Let us now analyze the behavior of the various excitation gaps.  
Following previous treatment of the SPT phase,~\cite{EF15,LEF15} 
we define the spin-, two-particle charge-, and neutral gaps as 
\begin{eqnarray}
 \Delta_{\rm s}&=&E_0(N,1)-E_0(N,0)\, , \\
 \Delta_{\rm c}&=&\frac{1}{2}\{E_0(N+2,0)+E_0(N-2,0)-2E_0(N,0)\}\, , 
 \nonumber\\
\end{eqnarray}
and  
\begin{eqnarray}
 \Delta_{\rm n}=E_1(N,0)-E_0(N,0)\, ,
\end{eqnarray}
respectively.
Here, $E_0(N_{\rm e}, S_{\rm tot}^z)$ denotes the ground-state 
energy of the finite system with $L$ sites, given the number of 
electrons $N_{\rm e}$ and the  $z$-component of total spin $S_{\rm tot}^z$.
$E_1(N_{\rm e},S_{\rm tot}^z)$ is the corresponding energy of the
first excited state. 

In the pure EHM ($\delta=0$), at small--to--intermediate $U/t$ and $V/t$, 
both $\Delta_{\rm c}$ and $\Delta_{\rm n}$ vanish at the BOW-CDW
transition, whereas $\Delta_{\rm s}$ stays finite. 
Turning on the dimerization $\delta$, also the
charge gap becomes finite, while the neutral gap still closes linearly, 
reflecting the fact that the transition point belongs to the Ising
universality class, see Fig.~\ref{gaps}(a) for $U/t=4$, where $V_{\rm c}/t\simeq2.503$.

By contrast, in the strong-coupling regime, the neutral gap stays 
finite passing the transition point, see Fig.~\ref{gaps}(b) for $U/t=12$.
Most strikingly, the spin gap exhibits a jump at the transition point
($V_{\rm c}/t\simeq6.192$), which indicates a first-order transition.

\subsection{Entanglement entropy}
\label{sec-ent}

\begin{figure}[t]
 \includegraphics[width=\columnwidth]{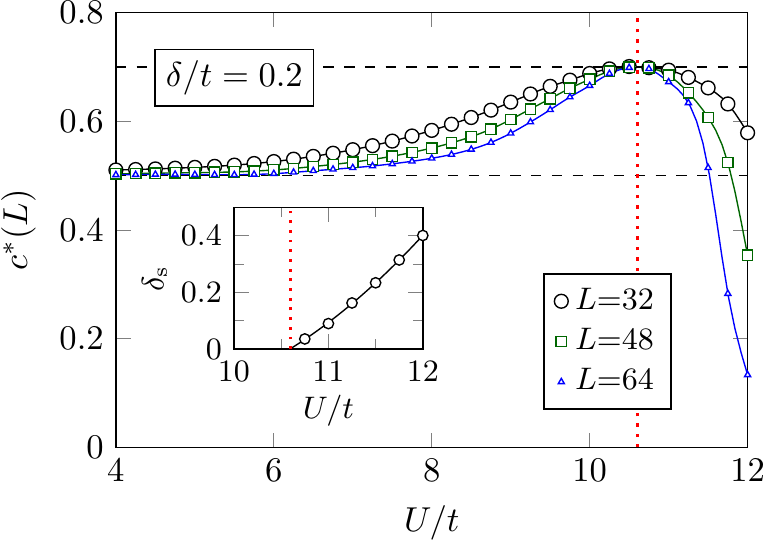}
 \caption{(Color online) Central charge $c^\ast(L)$ 
 along the PI-CDW transition line for $\delta/t=0.2$. 
 DMRG data (obtained with PBC)
 indicate the Ising universality class ($c=1/2$) 
 for $U<U_{\rm t}$ and, most notably, a tricritical Ising point 
 with $c=7/10$ at $U_{\rm t}$ (red dotted line).
 (Inset) Jump-value of the spin gap for $U\gtrsim U_{\rm t}$. The infinite MPS 
 data 
 point to a first order transition. 
 }
 \label{susy}
\end{figure}
We finally determine the universality class of the PI-CDW quantum phase 
transition. When the system becomes critical, the central charge $c$
can easily be deduced from the entanglement entropy.~\cite{ELF14,EF15}
CFT tells us that the von Neumann entropy 
for a system with PBC is\cite{CC04} 
\begin{eqnarray}
 S_L(\ell)=\frac{c}{3}\ln
  \left[  \frac{L}{\pi}\sin\left(\frac{\pi\ell}{L}\right)
  \right]
  +s_1\;,
\end{eqnarray}
where $s_1$ is a non-universal constant. In the face of the doubled 
unit cell of the SPT phase the related formula for the central charge
should be modified as~\cite{Ni11} 
\begin{eqnarray}
 c^\ast(L) \equiv \frac{3[S_L(L/2-2)-S_L(L/2)]}{\ln\{\cos[\pi/(L/2)]\}}\;.
\label{cstar}
\end{eqnarray}

Figure~\ref{susy} displays $c^\ast(L)$ along the PI-CDW transition line, varying $U$ and $V$ 
simultaneously at fixed dimerization strength $\delta/t=0.2$.
With increasing $U$, we find clear evidence for a crossover from 
$c^\ast(L)\simeq1/2$ to $c^\ast(L)\simeq7/10$, 
which signals Ising tricriticality.

Alternatively, the tricritical Ising point can be estimated 
from the magnitude of the jump of the spin gap,
$\delta_{\rm s}$,
see inset of Fig.~\ref{gaps} for $U/t=12$.
$\delta_{\rm s}$ should be finite for $U>U_{\rm t}$, and 
is expected to vanish at the tricritical Ising point, where $U=U_{\rm t}$. 
This is confirmed by the inset of Fig.~\ref{susy}. Obviously, 
$\delta_{\rm s}$ closes at $U_{\rm t}/t\simeq10.6$, in accord  with the 
critical value estimated from the numerically obtained central charge $c^\ast(L)$
in the main panel.

\section{Field theory analysis}
\label{sec:FT}
The weak-coupling regime $U$, $V\ll t$ of the model~\eqref{EPHM}
can be analyzed by field theory methods.\cite{TF04,BEG06} A standard
bosonization analysis~\cite{GNT99,EFGKK05} leads to the following form
of the low-energy Hamiltonian: 
\begin{eqnarray}
 {\cal H}&=&\sum_{\alpha=\textrm{c},\textrm{s}}\frac{v_\alpha}{16\pi}
  \left[(\partial_x\Phi_\alpha)^2+(\partial_x\Theta_\alpha)^2\right]
  +{\cal H}_{\textrm{int}}\, ,
  \nonumber \\
 {\cal H}_{\textrm{int}}&=&
  -\lambda_{\textrm{c}}\cos(\beta_{\textrm{c}}\Phi_{\textrm{c}})
  \nonumber\\
  &&+\lambda_{\textrm{s}}
   \left[
    \cos\left(\Phi_{\textrm{s}}\right)
    +\frac{a_0^2}{16}
    \left[(\partial_x\Theta_{\textrm{s}})^2
     -(\partial_x\Phi_{\textrm{s}})^2
    \right]
   \right]
  \nonumber\\
  &&+\lambda_{\delta}\cos\left(\frac{\Phi_{\textrm{s}}}{2}\right)
   \cos\left(\frac{\beta_{\textrm{c}}}{2}\Phi_{\textrm{c}}\right)
  \nonumber\\
  &&+\lambda_{\delta}^\prime\cos\left(\frac{\Phi_{\textrm{s}}}{2}\right)
   \cos\left(\frac{3\beta_{\textrm{c}}}{2}\Phi_{\textrm{c}}\right)
   +\dots \,.
\label{HQFT}
\end{eqnarray}
Here, $a_0$ is the lattice spacing, $\Phi_{\rm s,c}$ are canonical Bose fields
associated with the collective spin and charge degrees of freedom, and
$\Theta_{\rm s,c}$ the associated dual fields fulfilling
\begin{equation}
[\Phi_\alpha(x),\Theta_{\alpha^\prime}(x^\prime)]=4\pi
i\delta_{\alpha,\alpha^\prime}
\textrm{sgn}(x-x^\prime).
\end{equation}
The parameters $\beta_{\rm c}$, $\lambda_{\rm c,s}$, $\lambda_{\delta}$,
$\lambda'_{\delta}$, $v_{\rm c,s}$ can be determined at weak coupling
$U,V,\delta\ll t$. Compared to Ref.~[\onlinecite{BEG06}] we have retained one
higher harmonic in the interaction potential between spin and
charge degrees of freedom. The reason for this will become clear later
on. 
\subsection{Quantum critical line}
It was shown in  Refs.~[\onlinecite{TF04}] and [\onlinecite{BEG06}] 
that for appropriate choices of the parameters $U$, $V$, and $\delta$ 
the spin sector is gapped, while the charge sector undergoes 
a quantum phase transition. 
In the vicinity of this critical line we have
\begin{eqnarray}
 \cos\left(\frac{\Phi_{\textrm{s}}}{2}\right)\neq0\, .
\end{eqnarray}
Integrating out the massive spin degrees of freedom then
leads to an effective low-energy description of the charge
sector by a triple sine-Gordon model
\begin{eqnarray}
 {\cal H}_{\textrm{c}}^{\textrm{eff}}
  &=& \frac{v}{16\pi}
  \left[(\partial_x\Phi_{\textrm{c}})^2
   +(\partial_x\Theta_{\textrm{c}})^2
  \right]
  +g_\delta\cos\left(\frac{\beta_{\textrm{c}}}{2}\Phi_{\textrm{c}}\right)
  \nonumber \\
 &&+g_{\textrm{c}}\cos\left(\beta_{\textrm{c}}\Phi_{\textrm{c}}\right)
  +g_\delta^{\prime}
  \cos\left(\frac{3\beta_{\textrm{c}}}{2}\Phi_{\textrm{c}}\right)
  +\dots\, .
\label{Hc}
\end{eqnarray}
If we neglect the last term, we arrive at the two-frequency
sine-Gordon model discussed in Ref.~[\onlinecite{BEG06}]. It exhibits a
quantum phase transition in the Ising universality class.~\cite{DM98}
In the classical limit $\beta_{\textrm c}\to 0$, this corresponds to values of
$g_{\textrm{c}}$ and $g_\delta$ such that the quadratic terms in the
expansion of the cosines precisely cancel.
The reason for retaining the last term in (\ref{Hc}) is now clear: by
fine-tuning the parameters $g_{\rm c}$, $g_\delta$, $g'_\delta$ in the
classical limit, we can set the coefficient of the quartic term in
the expansion of the interaction potential to zero as well,  
which corresponds to a phase transition in the tricritical Ising
universality class. This scenario is known to persist in the full
quantum theory.~\cite{To04}

It is important to note that while the field theories (\ref{HQFT}) and
(\ref{Hc}) are initially derived in the limit $U,V,\delta\ll t$, they
have a wider regime of applicability, provided that their parameters
are adjusted appropriately. In the following we will assume that the
description (\ref{Hc}) applies along the line of quantum phase
transitions even at large values of $U/t$ and $V/t$. This will allow 
us to make predictions for the large distance behavior of various
correlation functions, which then can be tested by numerical
computations for the lattice model.

\subsection{Density correlations}
In the field theory limit, the bosonized form of the 
electron density is
\begin{eqnarray}
 n_j\to\rho_0(x)+(-1)^j\rho_\pi(x)\ ,\quad
  x=ja_0\, ,
\end{eqnarray}
where 
\begin{eqnarray}
 \rho_0(x)&=&{\rm const}-\frac{\beta_{\rm c}}{2\pi}\partial_x\Phi_{\textrm{c}}+
\hat{A}_0\partial_x\Phi_{\rm c}\cos\big(\frac{\Phi_{\rm s}}{2}\big)
 +\dots\, ,
  \nonumber\\
 \rho_\pi(x)&=&\hat{A}_\pi\sin\left(
		       \frac{\beta_{\textrm{c}}}{2}\Phi_{\textrm{c}}
		      \right)
                  \cos\left(
		       \frac{\Phi_{\textrm{s}}}{2}
		      \right)+\dots\, .
\end{eqnarray}
Here we have absorbed Klein factors into the non-universal amplitudes
$\hat{A}_{0,\pi}$. Importantly, at half-filling the smooth component $\rho_0(x)$
does not contain a $4k_{\rm F}$ umklapp contribution.\cite{CE02} 
As this is quite important, it is worthwhile to review the
derivation of this fact. We note that the Hamiltonian~\eqref{EPHM} is
invariant under the particle-hole transformation 
\begin{eqnarray}
 \hat{C}\hat{c}_{j,\sigma}\hat{C}^\dagger=(-1)^j\hat{c}_{j,-\sigma}.
\label{particle-hole-sym}
\end{eqnarray}
The electron density operator is odd under~\eqref{particle-hole-sym} 
\begin{eqnarray}
 \hat{C}(\hat{n}_j-1)\hat{C}^\dagger=1-\hat{n}_j\, .
 \label{symmetry}
\end{eqnarray} 
In the field theory Eq.~\eqref{particle-hole-sym} is implemented as follows
\begin{eqnarray}
 \hat{C}\varphi_{\textrm{c}}\hat{C}^\dagger&=&-\varphi_{\textrm{c}}\, , \ \ \
 \hat{C}\bar{\varphi}_{\textrm{c}}\hat{C}^\dagger=-\bar{\varphi}_{\textrm{c}}\, , 
 \nonumber \\
 \hat{C}\varphi_{\textrm{s}}\hat{C}^\dagger&=&\varphi_{\textrm{s}}\, , \ \ \
 \hat{C}\bar{\varphi}_{\textrm{s}}\hat{C}^\dagger=\bar{\varphi}_{\textrm{s}}\, , 
 \nonumber \\
 \hat{C}\eta_{\sigma}\hat{C}^\dagger&=&\eta_{-\sigma}\, , \ \ \
 \hat{C}\bar{\eta}_{\sigma}\hat{C}^\dagger=\bar{\eta}_{-\sigma}\, . 
\label{trafo}
\end{eqnarray}
Here $\eta_\uparrow$, $\eta_\downarrow$, $\bar{\eta}_\downarrow$, and 
$\bar{\eta}_\uparrow$ are Klein factors, \emph{cf.} Ref.~[\onlinecite{EPS15}].
At general band filling, the $4k_{\textrm{F}}$-term in the charge
density takes the form
\begin{eqnarray}
 \rho_{4k_{\textrm F}}(x)=
  A_{4k_{\textrm F}}\eta_\uparrow{\bar \eta}_\uparrow
                    \eta_\downarrow{\bar \eta}_\downarrow
		    \cos(\beta_{\textrm{c}}\Phi_{\textrm{c}}-4k_{\textrm F}x)
		    +\dots\, .
\end{eqnarray}
Eq.~(\ref{trafo}) implies that at half-filling ($4k_{\textrm{F}}x=0
\mod 2\pi$) we have
\begin{eqnarray}
 C\rho_{4k_{\textrm F}}(x)C^\dagger=\rho_{4k_{\textrm F}}(x)\, ,
\end{eqnarray}
which can be reconciled with Eq.~\eqref{symmetry} only by taking
$A_{4k_{\textrm F}}=0$.

In the vicinity of the quantum critical line, we can again
integrate out the gapped spin degrees of freedom and arrive at
\begin{eqnarray}
\rho_0(x)&=&{\rm const}+B_0\partial_x\Phi_{\textrm{c}}+\dots\ ,\nonumber\\
\rho_\pi(x)&=&B_\pi\sin\left(\frac{\beta_{\textrm{c}}}{2}\Phi_{\textrm{c}}\right)
 +\dots\, .
\label{rho0pi}
\end{eqnarray}
Finally, we need to relate our charge boson to the primary
fields in the tricritical Ising model. This can be done 
by referring to the Landau-Ginzburg description of the
transition, see, e.g., Ref.~[\onlinecite{LMC91}]. Expanding our 
low-energy effective theory (\ref{Hc}) for $\beta_{\rm c}\ll 1$,
we obtain the Landau-Ginzburg model
\begin{equation}
{\cal L}\sim\frac{v}{16\pi}
\Phi_{\rm c}\left(\partial_x^2-\frac{\partial_t^2}{v^2}\right)\Phi_{\rm c}
-\lambda_2\Phi_{\rm c}^2-\lambda_4\Phi_{\rm c}^4-\lambda_6\Phi_{\rm c}^6
 +\dots .
\end{equation}
In this limit, we can then use Ref.~[\onlinecite{LMC91}] to relate
local operators in our theory to primary fields in the TIM. In
particular, one has 
\begin{eqnarray}
\Phi_{\rm c}(x)&\leftrightarrow& \sigma(x) \ ,\nonumber\\
:\Phi^2_{\rm c}(x):&\leftrightarrow& \epsilon(x) \ ,\nonumber\\
:\Phi^3_{\rm c}(x):&\leftrightarrow& \sigma'(x) \ ,\nonumber\\
:\Phi^4_{\rm c}(x):&\leftrightarrow& \epsilon'(x)\, ,
\end{eqnarray}
where $\sigma$, $\epsilon$, $\sigma'$, and $\epsilon'$ are respectively
the magnetization field, energy density, sub-magnetization, and vacancy
density in the TIM. Proceeding in the same way for the components of
the charge density (\ref{rho0pi}) then suggests the following
identifications: 
\begin{eqnarray}
 \rho_\pi(x) &\sim& A \sigma(x)+\dots\, ,
 \nonumber\\
 \rho_0(x) &\sim& {\rm const}+Ba_0\partial_x\sigma(x)+\dots\, . 
\label{rhocontent}
\end{eqnarray}
Using the known results for correlation functions in the TIM, we
then arrive at the following prediction for the density-density
correlator at the Ising tricritical point: 
\begin{eqnarray}
 \langle(\hat{n}_{j+\ell}-1)(\hat{n}_j-1)\rangle\sim
  (-1)^\ell\frac{A^2}{\ell^{3/20}}+\dots\, ,\ \ell\gg 1.
  \label{nn-corr-tri}
\end{eqnarray}
We may isolate the subleading behavior by considering smooth and
staggered combinations of the density on the lattice:
\begin{eqnarray}
 \hat{n}_j^{\rm st}&=&(-1)^j(\hat{n}_{j}-\hat{n}_{j+1})\sim 2A\sigma(x)+\dots\, ,
  \nonumber\\
 \hat{n}_j^{\rm
   sm}&=&\frac{\hat{n}_{j}+\hat{n}_{j+1}}{2}-1\sim(B-(-1)^jA)
a_0\partial_x\sigma+\dots\, .
 \nonumber\\
\label{local-n-stsm}
\end{eqnarray}
The TIM predictions for two point functions of these operators are
\begin{eqnarray}
 \langle\hat{n}_{j+\ell}^{\rm st}\hat{n}_j^{\rm st}\rangle
  &\sim& 4A^2\ell^{-3/20}+\dots\, ,
  \label{nst-corr-tri}
 \\
 \langle\hat{n}_{j+\ell}^{\rm sm}\hat{n}_j^{\rm sm}\rangle
  &\sim& C_{j,\ell}\ell^{-43/20}+\ldots\ ,\nonumber\\
C_{j,\ell}&=&-\frac{69}{400}
\begin{cases}
B^2-A^2/4 &  \ell \text{ odd}\\
[B-(-1)^jA/2]^2 & \ell \text{ even.}
\end{cases}
  \label{nsm-corr-tri}
\end{eqnarray}

\begin{figure}[tb]
 \includegraphics[width=\columnwidth]{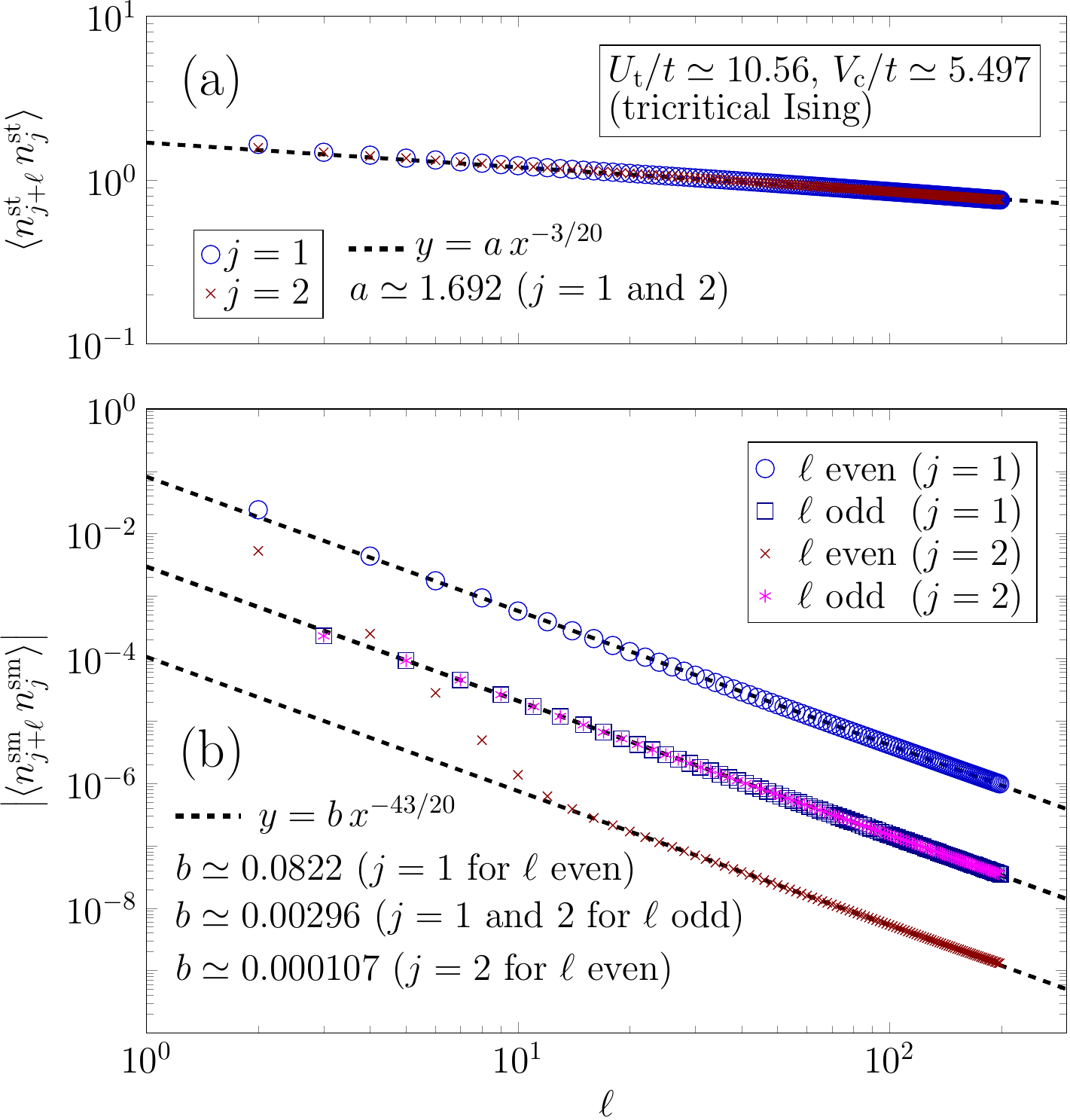}
 \caption{(Color online) Density-density correlation functions
 at the tricritical Ising point for $\delta/t=0.2$. Data obtained 
 by iDMRG with $\chi=1600$. The correlation functions  (symbols) show a power-law decay, in accordance 
 with the field theory predictions, 
 Eqs.~\eqref{nst-corr-tri} and \eqref{nsm-corr-tri}.
 }
 \label{dens-dens-tri}
\end{figure}
The predictions \eqref{nst-corr-tri} and \eqref{nsm-corr-tri} 
can now be compared with iDMRG simulations of the 1D lattice
model~\eqref{EPHM}. Figure~\ref{dens-dens-tri} shows the iDMRG 
results for two point functions of the (a) staggered and (b) smooth
combinations of the particle density at the TIM critical point of the
lattice model. The results for $\langle\hat{n}_{j+\ell}^{\rm
  st}\hat{n}_j^{\rm st}\rangle$ are seen to be in excellent agreement
with the leading $\ell^{-3/20}$ dependence at long distances predicted
by Eq.~\eqref{nst-corr-tri} for both $j=1$ and $j=2$.
To test the second prediction in Eq.~\eqref{nsm-corr-tri}, we consider
separately the cases of even and odd $\ell$  for $j=1$ and $j=2$, and
plot the absolute value of  $\langle\hat{n}_{\ell+1}^{\rm
  sm}\hat{n}_1^{\rm sm}\rangle$ in Fig.~\ref{dens-dens-tri}(b).
Again the numerical data are seen to be in excellent agreement with the
predicted $\ell^{-43/20}$ dependence at large separations. The
prefactors for the power laws extracted from our iDMRG data are in
very good agreement with the prediction of Eq.~\eqref{nsm-corr-tri} as
well. 

\subsection{BOW correlations}
\begin{figure}[t]
 \includegraphics[width=\columnwidth]{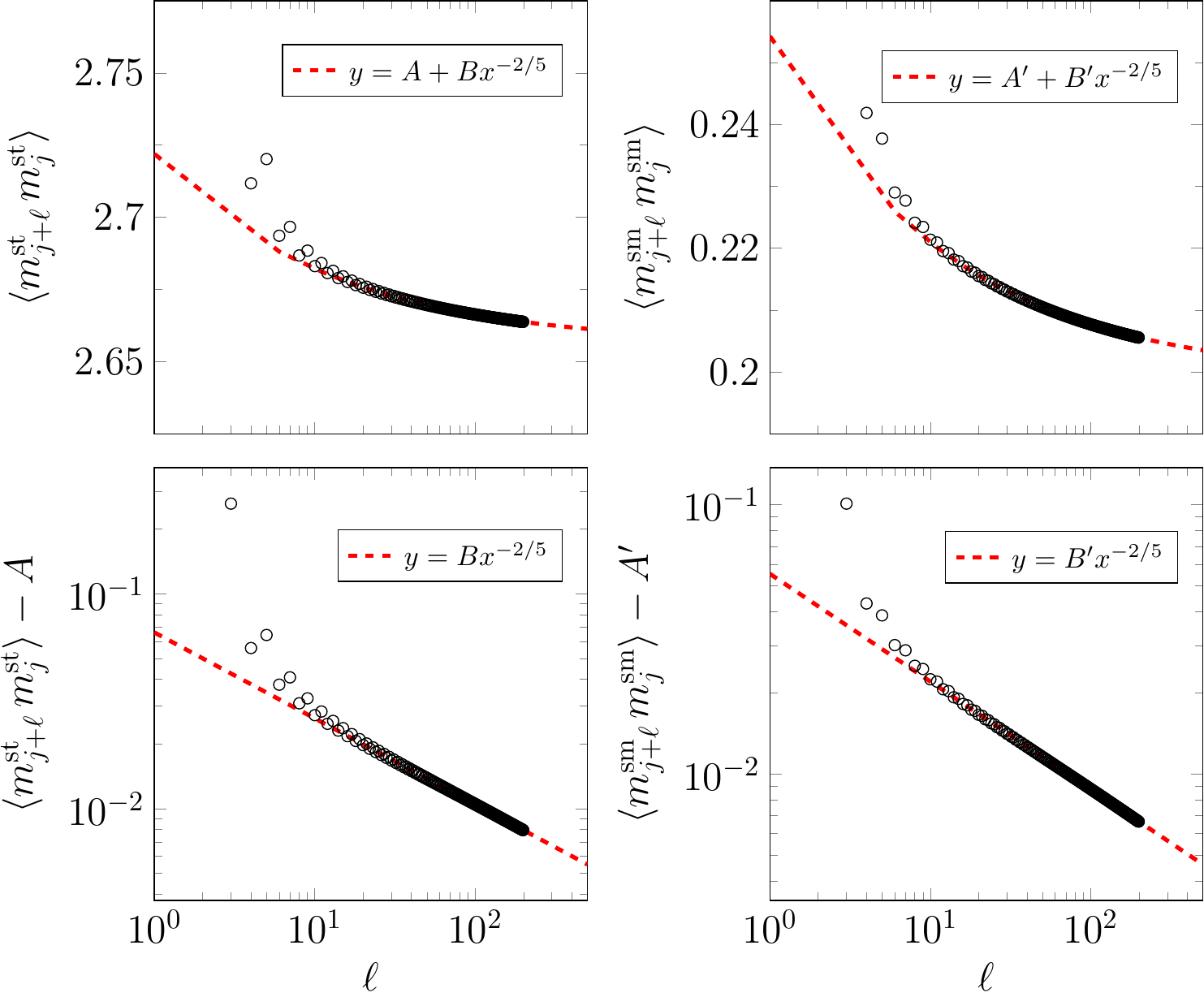}
 \caption{(Color online) BOW correlation functions
at the tricritical Ising point for $\delta/t=0.2$ computed by
iDMRG with $\chi=1600$. (Top) The asymptotic values
for the two-point functions of staggered and smooth combinations of
the BOW density are estimated by fitting to Eqs.~\eqref{mst-corr-tri}
and \eqref{msm-corr-tri}. (Bottom) log-log plots of the same
correlation functions with the asymptotic values subtracted show
power-law decay compatible with Ising tricriticality.
 }
 \label{BOW-corr}
\end{figure}
The BOW order parameter is given by 
$\hat{m}_{\rm BOW}=(1/L)\sum_j \hat{m}_j$
with
\begin{eqnarray}
 \hat{m}_j=(-1)^j\sum_\sigma
  \left[
   \hat{c}_{j\sigma}^{\dagger}\hat{c}_{j+1\sigma}^{\phantom{\dagger}}
   +{\rm h.c.}
  \right]\,.
\end{eqnarray}
The BOW order parameter is always non-zero in the vicinity of the
transition 
\begin{equation}
\langle\hat{m}_{\rm BOW}\rangle\neq 0.
\end{equation}
The bosonized expression for $\hat{m}_{\rm BOW}$ is
\begin{eqnarray}
\hat{m}_j&\sim&(-1)^j{\rm const}+
\hat{C}_\pi\cos\left(
		       \frac{\beta_{\textrm{c}}}{2}\Phi_{\textrm{c}}
		      \right)
                  \cos\left(
		       \frac{\Phi_{\textrm{s}}}{2}
		      \right)\nonumber\\
&&+(-1)^j\hat{C}_0\cos\big(\beta_{\rm c}\Phi_{\rm c}\big)+\dots .
\end{eqnarray}
We now proceed in the same way as for the charge density. We integrate
out the gapped spin degrees of freedom, then expand for small
$\beta_{\rm c}$, and finally use the Landau-Ginzburg description to identify
which operators in the TIM dominate the long distance behavior of the
BOW correlations. The main difference compared to the charge density
is that the BOW order parameter is even under charge conjugation, and
concomitantly we find
\begin{eqnarray}
\hat{m}_j&\sim& \langle\hat{m}_{\rm BOW}\rangle+D_0\epsilon(x)\nonumber\\
&&+(-1)^j\left[D_1+D_2\epsilon(x)\right]+\ldots\ .
\end{eqnarray}
We again form smooth and staggered combinations,
\begin{eqnarray}
 \hat{m}_j^{\rm st}&=&(-1)^j(\hat{m}_{j}-\hat{m}_{j+1})
\sim 2 \left[D_1+D_2\epsilon(x)\right]+\dots\, ,
  \nonumber \\
 \hat{m}_j^{\rm sm}&=&\frac{\hat{m}_{j}+\hat{m}_{j+1}}{2}\sim
\langle\hat{m}_{\rm BOW}\rangle+D_0\epsilon(x)+\dots\, .
\label{bowcontent}
\end{eqnarray}
The TIM predictions for BOW correlations are then
\begin{eqnarray}
 \langle\hat{m}_{j+\ell}^{\rm st}\hat{m}_j^{\rm st}\rangle
  &\sim& 4\left[D_1^2+D_2^2\ell^{-2/5}\right]
  +\dots\, ,
  \label{mst-corr-tri}
  \\
 \langle\hat{m}_{j+\ell}^{\rm sm}\hat{m}_j^{\rm sm}\rangle
  &\sim& \langle\hat{m}_{\rm BOW}\rangle^2+D_0^2\ell^{-2/5}
  +\dots\, .
  \label{msm-corr-tri}
\end{eqnarray}


These predictions can be compared to iDMRG computations in
Fig.~\ref{BOW-corr}. In order to remove the constant terms  
in Eqs.~\eqref{mst-corr-tri} and \eqref{msm-corr-tri}, we first fit 
the numerical results to the functional form $y=A+Bx^{-2/5}$. This
allows us to extract the constants as shown in the upper panels in
Fig.~\ref{BOW-corr}. Subtracting the estimated constants from original
data, both staggered and smooth correlation functions are seen to
decay in a power-law fashion compatible with the TIM prediction.

\subsection{Spin correlations}
\begin{figure}[t]
 \includegraphics[width=\columnwidth]{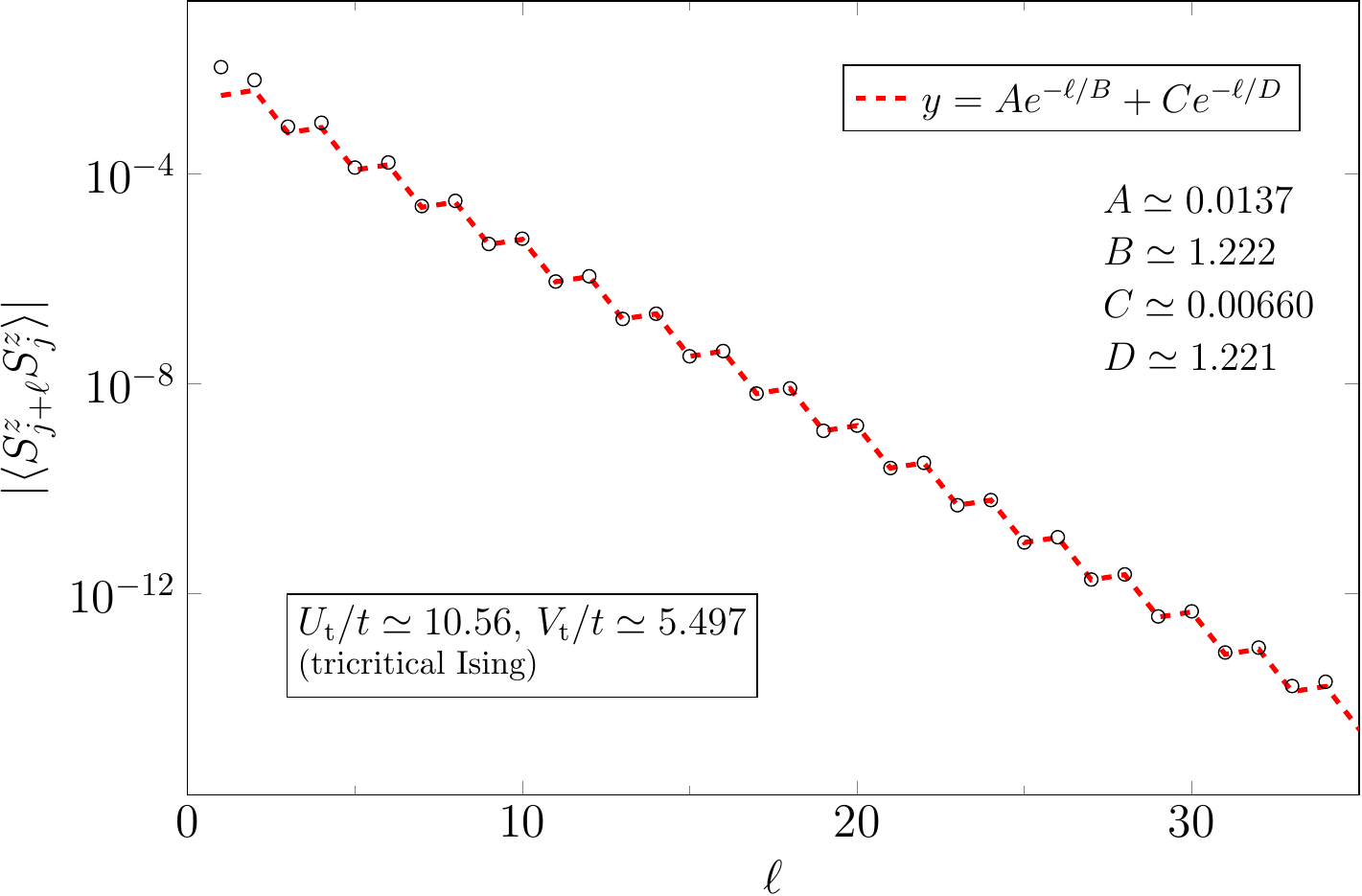}
 \caption{(Color online) Spin correlation function (symbols)
 at the tricritical Ising point for $\delta/t=0.2$ using
 the iDMRG with $\chi=1600$, showing exponential decay.
 The line is a fit to Eq.~\eqref{szsz-corr}.
 }
 \label{SzSz-corr}
\end{figure}

As the spin sector is gapped, we expect an exponential decay
for the spin two-point function
\begin{eqnarray}
 \langle \hat{S}^z_{j+\ell}\hat{S}^z_j\rangle \sim
E_0 e^{-\ell/\xi_1}+E_1 (-1)^\ell e^{-\ell/\xi_2}\, .
\label{szsz-corr}
\end{eqnarray}
Here we have used that the low energy degrees of freedom in the spin
sector occur at wave numbers zero and $\pi$. 
This behavior is again in good agreement with iDMRG computations
as shown in Fig.~\ref{SzSz-corr}. 
The correlation lengths extracted by fitting the iDMRG results to
Eq.~\eqref{szsz-corr} are found to be in reasonable agreement with
the corresponding eigenvalue of the transfer matrix 
$\xi_1\simeq1.225$.

\vspace*{0.5cm}

To summarize this section, we have seen that field theory
predictions obtained by means of a triple sine-Gordon model
description of the tricritical Ising transition are in excellent
agreement with iDMRG computations for the lattice model. This
firmly establishes that the critical endpoint is in the universality
class of the TIM. We note that an analogous field theory description
applies along the entire Ising critical line. Here, field theory
predictions are again in excellent agreement with iDMRG computations
as shown in Appendix.

\section{Conclusions}
\label{conclusion}
We have revisited ground-state properties of the one-dimensional
half-filled extended Hubbard model with staggered bond
dimerization. We have employed a combination of numerical and
analytical techniques to map out the ground-state phase diagram in
detail, and identify all quantum critical regions. At fixed
dimerization $\delta$, there are two distinct phases. A CDW phase at
large $V\agt U$ is separated from a PI phase 
at $U\agt V$ by an Ising critical line, that terminates 
in a critical point which we have
shown to be in the universality class of the tricritical Ising model.
Our identification was based on a detailed analysis of both
entanglement entropy scaling and critical exponents describing the
power-law decay of several two-point correlation functions.

Correlation functions of local operators in the EHM with bond
dimerization access only the bosonic sector of the TIM CFT. 
This precludes us from directly
investigating the emergence of supersymmetry at low energies/long
distances. To ``see'' the fermionic sector one presumably would have
to consider correlation functions of suitably constructed non-local
operators. It would be interesting to investigate this possibility
further. Another issue worth pursuing is to investigate the scaling
regime around the TIM critical point in the framework of the EHM
with bond dimerization. 
It would be interesting to investigate whether it is possible to make
contact with the field theory predictions of Ref.~[\onlinecite{LMT08}].

\section*{Acknowledgments}
We thank P. Fendley and G. Mussardo for useful discussions. The iDMRG
simulations were performed using the ITensor library.~\cite{ITensor}
This work was supported by Deutsche Forschungsgemeinschaft (Germany),
SFB 652, project B5, and by the EPSRC under grant EP/N01930X/1 (FHLE).

\appendix
\section{Correlation functions on the Ising critical line}
\label{FT-Ising}
The tricritical Ising model describes the end point of a critical
line of Ising transitions, \emph{cf.} Fig.~\ref{pd}. The Ising
critical line was previously investigated by DMRG methods in
Ref.~[\onlinecite{BEG06}] and the critical exponents were extracted by
considering the scaling of the order parameter and spectral gap in the
vicinity of the transition. In this appendix we complement these
results by examining 
the power law behavior of correlations functions at the transition, 
i.e. the same diagnostics we used in the main text to identify 
the TIM critical point.

\begin{figure}[tb]
\includegraphics[width=\columnwidth]{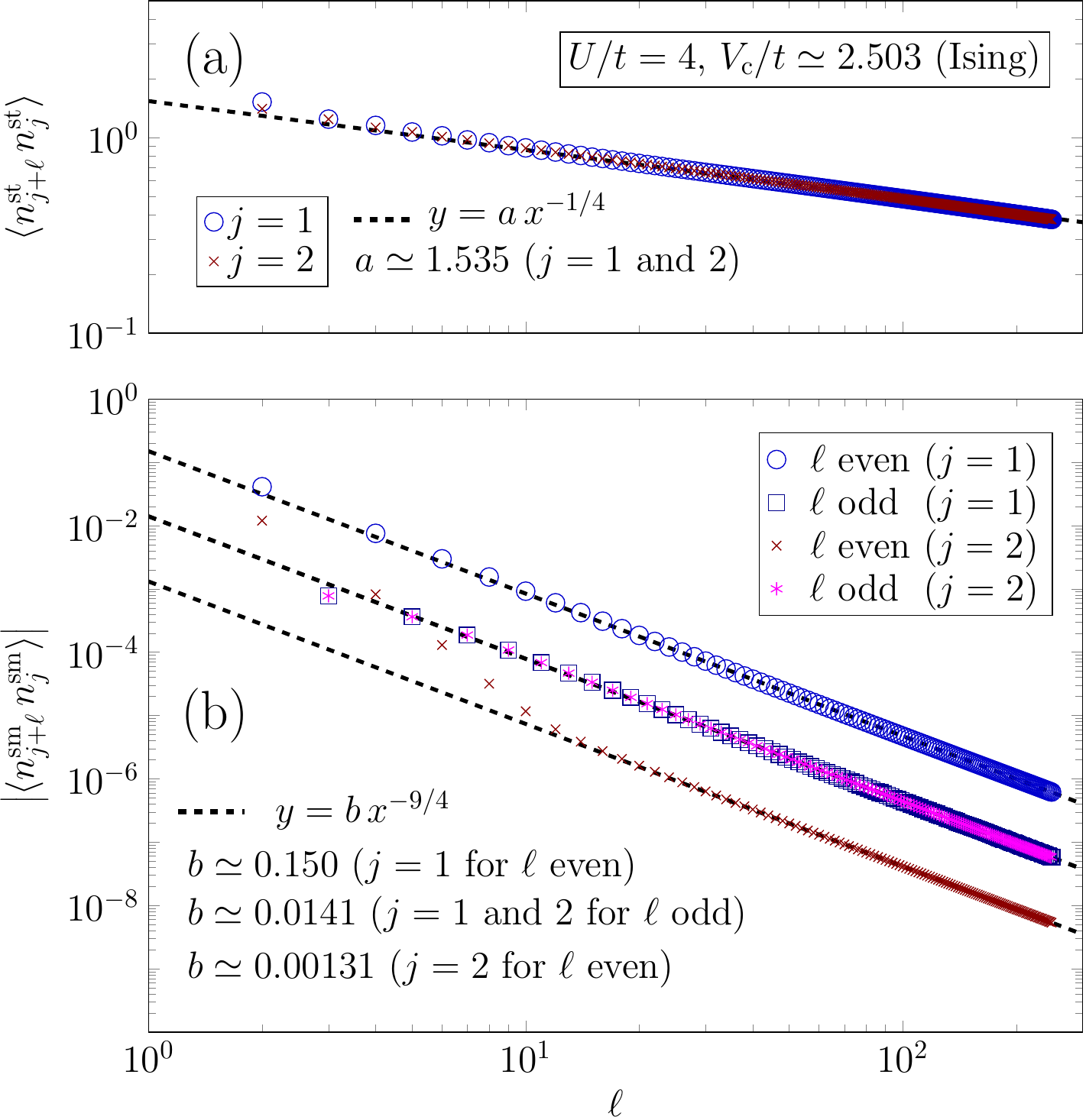}
\caption{(Color online) Density-density correlation functions at the
Ising transition point ($V_{\rm c}\simeq2.503$) for $U/t=4$ and
$\delta/t=0.2$, using the iDMRG with $\chi=1600$. 
(a) The correlator of the staggered combination is in excellent
agreement with Eq.~\eqref{nst-corr-ising} with $4\tilde{A}^2\approx1.535$.
(b) Correlations of the smooth combination $n^{\rm sm}_j$ are plotted
separately for odd and even $\ell$ with $j=1$ and 2. The data are in
excellent agreement with the prediction  Eq.~\eqref{nsm-corr-ising}.
 }
 \label{Ising-nst-nsm}
\end{figure}
The identification of operators is analogous to the TIM case. The
projections of the particle density and BOW order parameter onto
local fields in the Ising CFT are again of the form
(\ref{rhocontent}) and (\ref{bowcontent}), but $\sigma(x)$ and $\epsilon(x)$
are now the spin field and energy density of the Ising CFT. This leads
to the following prediction for the large distance asymptotics of the
density-density correlator
\begin{eqnarray}
  \langle(\hat{n}_{j+\ell}-1)(\hat{n}_j-1)\rangle 
  &\sim& (-1)^\ell\tilde{A}\ell^{-1/4}+\dots\, .
  \label{nn-corr-ising}
\end{eqnarray}
Considering smooth and staggered combinations defined in
\eqref{local-n-stsm} separately, we obtain
\begin{eqnarray}
 \langle\hat{n}_{j+\ell}^{\rm st}\hat{n}_j^{\rm st}\rangle
  &\sim& 4\tilde{A}^2\ell^{-1/4}+\dots\, ,
  \label{nst-corr-ising} \\
 \langle\hat{n}_{j+\ell}^{\rm sm}\hat{n}_j^{\rm sm}\rangle
  &\sim& \tilde{C}_{j,\ell}\ell^{-9/4}+\ldots\ ,\nonumber\\
\tilde{C}_{j,\ell}&=&-\frac{5}{16}
\begin{cases}
\tilde{B}^2-\tilde{A}^2/4 &  \ell \text{ odd}\\
[\tilde{B}-(-1)^j\tilde{A}/2]^2 & \ell \text{ even.}
\end{cases}
  \label{nsm-corr-ising}
\end{eqnarray}

These predictions are in excellent agreement with iDMRG computations
for the lattice model on the Ising critical line as is shown in
Fig.~\ref{Ising-nst-nsm}. 

\begin{figure}[b]
 \includegraphics[width=\columnwidth]{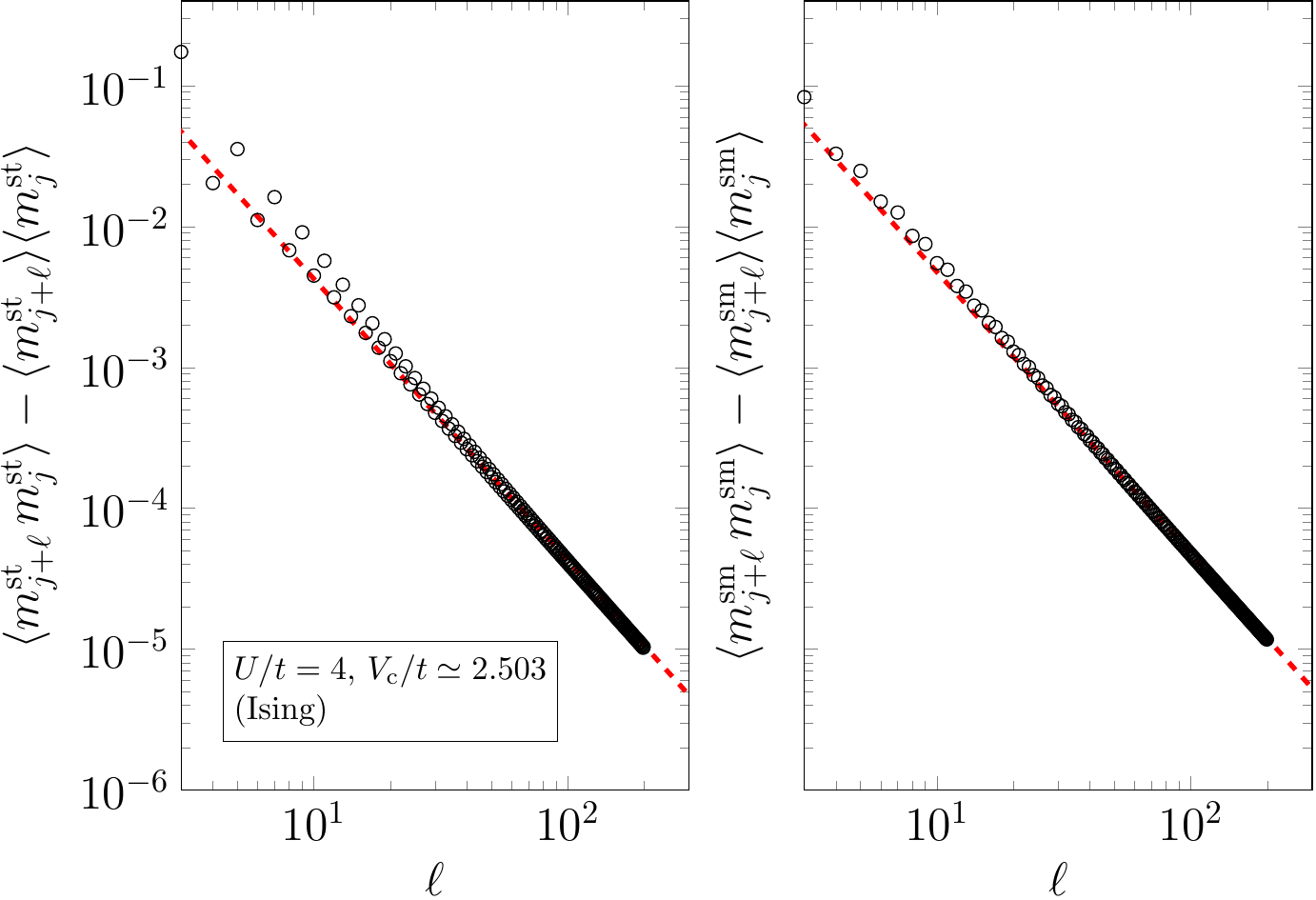}
 \caption{(Color online) BOW correlations 
 at the Ising transition point 
 for $U/t=4$ and $\delta/t=0.2$. The correlators exhibit a power-law
 decay consistent with the field theory predictions,
 Eqs.~\eqref{msm-corr-ising}. 
 }
 \label{Ising-mst-msm}
\end{figure}

The field theory predictions for staggered and smooth combinations of
the BOW order parameter on the Ising transition line are
\begin{eqnarray}
 \langle\hat{m}_{j+\ell}^{\rm st}\hat{m}_j^{\rm st}\rangle
  &\sim& (-1)^\ell\left[\tilde{C}_4^2+\tilde{C}_5\ell^{-2}\right]
  +\dots\, ,
  \nonumber \\
 \langle\hat{m}_{j+\ell}^{\rm sm}\hat{m}_j^{\rm sm}\rangle
  &\sim& \langle\hat{m}_{\rm BOW}\rangle^2+\tilde{C}_6\ell^{-2}
  +\dots\, .
  \label{msm-corr-ising}
\end{eqnarray}
We can remove the constant contributions by considering connected
correlators, which in turn exhibit power-law decay to zero at large
distances. The iDMRG results shown in Fig.~\ref{Ising-mst-msm} agree
perfectly with the predicted $\ell^{-2}$ power-law decay.
As a consistency check we have extracted the value of
$\langle\hat{m}_{\rm BOW}\rangle$ by fitting the long-distance
behavior of two-point function of $\hat{m}_j^{\rm sm}$ to the form
(\ref{msm-corr-ising}). We find it to be in excellent agreement with
the value obtained by computing the one-point function.

We note that the agreement between our numerical data and field
theory predictions is much better along the Ising transition line
that at the TIM critical point. There are two reasons for this. First,
at fixed $U/t$, the Ising transition point ($V_{\rm c}/t$) can be
determined more accurately than the location of the TIM transition,
where two parameters ($U$ and $V$) have to be fine-tuned
simultaneously. Second, the corrections to scaling are different in
both cases.

%

\end{document}